# Hyperparameter tuning of optical neural network classifiers for high-order Gaussian beams

Shunsuke Watanabe,[1] Tomoyoshi Shimobaba,[1,*] Takashi Kakue,[1] and Tomoyoshi Ito[1]

[1]*Graduate School of Engineering, Chiba University, 1-33 Yayoi-cho, Inage-ku, Chiba 263-8522, Japan*
*\*shimobaba@faculty.chiba-u.jp*

**Abstract:** High-order Gaussian beams with multiple propagation modes have been studied for free-space optical communications. Fast classification of beams using a diffractive deep neural network (D$^2$NN) has been proposed. D$^2$NN optimization is important because it has numerous hyperparameters, such as interlayer distances and mode combinations. In this study, we classify Hermite–Gaussian beams, which are high-order Gaussian beams, using a D$^2$NN, and automatically tune one of its hyperparameters known as the interlayer distance. We used the tree-structured Parzen estimator, a hyperparameter auto-tuning algorithm, to search for the best model. Results indicated that classification accuracy obtained by auto-tuning hyperparameters was higher than that obtained by manually setting interlayer distances at equal intervals. In addition, we confirmed that accuracy by auto-tuning improves as the number of classification modes increases.

## 1. Introduction

Gaussian beams have a high-order propagation mode. Examples include Laguerre–Gaussian (LG) beams and Hermite–Gaussian (HG) beams, which have cylindrical symmetry and rectangular symmetry, respectively. LG beams are light with a helical wavefront known as optical vertex. LG beams have a quantity called orbital angular momentum of $l\hbar$ ($l$ represents an integer and $\hbar$ represents Planck's constant) per photon [1], which can increase the number of channels in free-space optical communications. An LG beam has orders $p$ and $l$ in the radial direction and azimuthal directions, respectively. $p$ represents an integer greater than or equal to zero, and $l$ represents a positive or negative integer; thus, theoretically an LG beam has an infinite number of propagation modes.

An HG beam is a rectangular beam with symmetry along the propagation axis and has an order $m$ in the x-direction and an order $n$ in the y-direction in the plane perpendicular to the direction of propagation z. Since both orders are integer values that are greater than or equal to zero, an HG beam also has an infinite number of propagation modes. Multiple-bit communication by assigning codes to each of these modes [2, 3] along with conventional amplitude shift keying (ASK) [4] and multiplexing using orthogonality [5] have been studied. Terabit free-space optcical communication with high-order Gaussian beams has already been achieved by combining coding and multiplexing [6].

A system for detecting the mode of a beam is required to realize multibit communications. Several methods for detecting a beam's code have been investigated, including mode sorters [7], computer-generated holograms displayed on a spatial light modulator (SLM) [8], and deep learning [6, 9]. Imaging using a charge coupled device (CCD) camera and computational load in neural networks are bottlenecks of these methods.

Beam classification using electrical deep neural networks based on semiconductors can be expected to have high accuracy; however, the classification speed is limited by the semiconductor's clock frequency, and power consumption is also high. Recently, an optical computer known as a diffractive deep neural network (D$^2$NN), which uses passive optical modulators, has been proposed [10]. In a D$^2$NN, training calculation is performed using a computer, and trained parameters are recorded in an optical modulator (photopolymer or SLM). Light modulators correspond to the layers of a neural network, and by arranging them in equal intervals, a deep neural network is mimicked. The input light is modulated in each layer, and a photodetector placed in the output layer detects the calculation result of D²NN. Since classification can be performed only by light propagation and diffraction, it is performed at the speed of light with a minute amount of power consumption.

Multimode classification of high-order Gaussian beams using a D²NN has been performed [6, 11, 12], as has multiplexed beam generation and separation [12, 13, 14]. In both cases, beam classification can be detected by a photodetector, which can speed up the detection speed by two to three orders of magnitude compared to a CCD camera.

However, in D²NN training, the number of layers, distance between layers, and choice of propagation mode are hyperparameters that should be optimized to improve classification accuracy. In general, hyperparameters are empirically determined.

In this study, we propose an automatic tuning of the hyperparameters of a D²NN for classifying HG beams of high-order Gaussian beams. Instead of manual tuning, automatic tuning algorithms using random and grid search and Bayesian optimization have been proposed to tune hyperparameters [15]. We use Bayesian optimization, tree-structured Parzen estimator (TPE), as an auto-tuning algorithm to optimize the interlayer distance and compare it with the case of equal spacing of layers. Phase disturbance due to the atmosphere was added to HG beams to assume free-space optical communications.

Section 2 describes HG beams and atmospheric turbulence, Section 3 explains the proposed method, Section 4 presents simulation results, and Section 5 concludes this study.

## 2. High-order Gaussian beam and atmospheric turbulence

In this session, we describe the HG beam used in this study and the disturbance model of the beam in the atmosphere.

### 2.1 Hermite–Gaussian beam

An HG beam is a beam having a rectangular symmetry [1], which has the following two degrees of freedom in the plane perpendicular to the optical axis: order $m$ in the x-direction and order $n$ in the y-direction. It is represented by

$$u_{m,n}^{HG}(x,y,z) = \frac{1}{w(z)} H_m\left(\frac{\sqrt{2}x}{w(z)}\right) H_n\left(\frac{\sqrt{2}y}{w(z)}\right) \exp(jkz) \\ \times \exp\left(-\frac{r^2}{w_0^2(1+z/z_R)}\right) \exp\left(j(1+m+n)\tan^{-1}\left(\frac{z}{z_R}\right)\right), \quad (1)$$

where $j = \sqrt{-1}$ $w_0$ represents the beam diameter at $z = 0$ and $H_n$ represents Hermite polynomials. Both $m$ and $n$ represent integers greater than or equal to zero, and there are theoretically infinite beam patterns. Figure 1 demonstrates examples of HG beams. As shown in Fig. 1, an HG beam is a rectangular beam with symmetry, and the number of divisions of the beam varies with $m$ and $n$ values. An HG beam has multiple propagation modes. In addition, multiple-bit communication is possible by assigning codes to each mode.

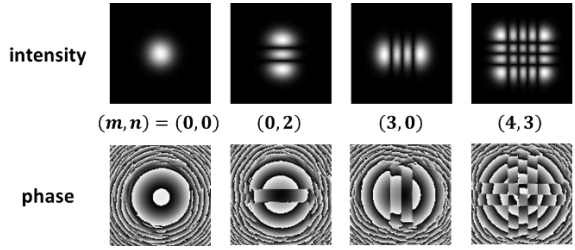

Figure 1：Examples of Hermite–Gaussian (HG) beams. ($m$, n) represents the orders of $m$ and $n$ for an HG beam, respectively.

## 2.2 Atmospheric turbulence

When light propagates through the atmosphere, the phase of the beam is disturbed. In this study, we used the von Karman model to add atmospheric disturbance to an HG beam [16]. The fluctuation spectrum of the refractive index $\phi_n(\kappa)$ is expressed by

$$\phi_n(\kappa) = 0.033 C_n^2 (\kappa^2 + k_0^2)^{-\frac{11}{6}} \exp\left(-\frac{\kappa^2}{k_m^2}\right), \tag{2}$$

where $k_0 = 2\pi/L_0$ and $k_m = 5.92/l_0$. $L_0$ is the external scale of the disturbance, and $l_0$ represents the internal scale of the disturbance. $\kappa$ represents the spatial frequency, and $C_n^2$ represents the atmospheric structure constant, which represents the strength of the disturbance caused by the atmosphere. By Markov approximation, the spectrum of the phase can be expressed by

$$\phi_\varphi(\kappa) = 2\pi k^2 z \phi_n(\kappa), \tag{3}$$

where $k$ represents the wavenumber of the beam and $z$ represents the propagation distance in the atmosphere. The phase disturbance of the propagated beam in the spatial domain can be expressed by

$$\varphi(x, y) = \mathrm{Re}\left\{F\left[C\left(\frac{2\pi}{N\Delta x}\right)\sqrt{\phi_\varphi(\kappa)}\right]\right\}, \tag{4}$$

where $\Delta x$ is the sampling pitch, $N$ represents the number of pixels in the height and width of the beam, and $C$ represents an $N \times N$ random complex matrix generated with mean 0 and variance 1. $F$ represents the Fourier transform and Re represents the operator to extract the real part. The resulting $\varphi$ represents the number of phases disturbed by the atmosphere.

## 3. Proposed method

Figure 2 shows the D²NN model used in this study. The D²NN consists of multiple diffraction layers. In the conventional method shown in Fig. 2(a), each layer is equally spaced, whereas, in the proposed method shown in Fig. 2(b), each layer is unequally spaced. Pixels in each layer correspond to neurons, and the light diffracted from the previous layer is optically coupled to each neuron in the next layer. The angular spectrum method was used for light propagation calculation. The lightwave $u^l$ propagating from the $l-1$ layer to $l$ layer is calculated by

$$u^l(x, y) = F^{-1}\{U^{l-1}(f_x, f_y) \cdot H(f_x, f_y)\}, \tag{5}$$

$$U^{l-1}(f_x, f_y) = F\{u^{l-1}(x, y) t^{l-1}(x, y)\}, \tag{6}$$

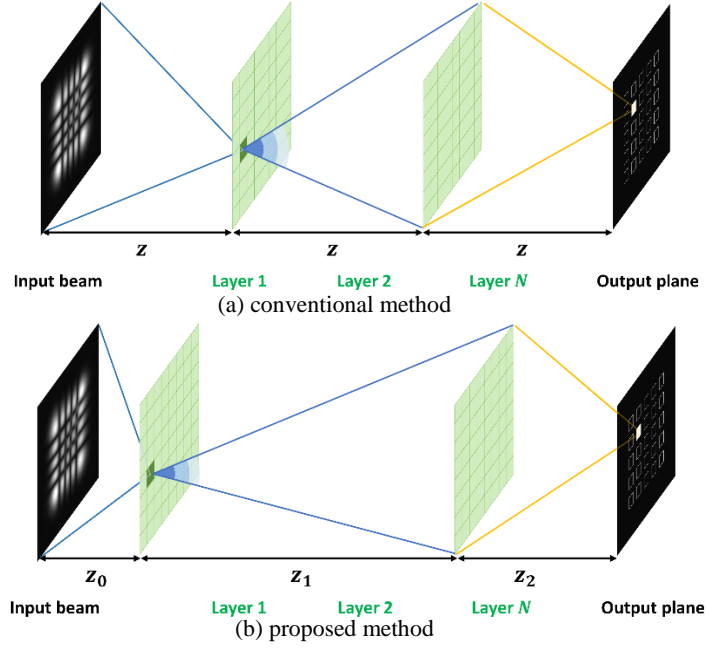

Figure 2：A model of diffractive deep neural network.

$$H(f_x, f_y) = \exp\left(j2\pi\Delta z\sqrt{\frac{1}{\lambda^2} - f_x^2 - f_y^2}\right). \quad (7)$$

where $f_x, f_y$ are the spatial frequencies, $F^{-1}$ represents the inverse Fourier transform, $\lambda$ is the wavelength of light, and $\Delta z$ is the propagation distance between adjacent layers. The transmission coefficient $t^l$ is the complex amplitude that changes the amplitude and phase of light passing through a neuron and is denoted by

$$t^l(x,y) = a^l(x,y)\exp(j\phi^l(x,y)), \quad (8)$$

where $a^l$ is the value of the amplitude and is a real number between 0 and 1. $\phi^l$ is the phase retardation, ranging from 0 to $2\pi$. In this study, we assume that layers in the D²NN are pure phase modulators with an amplitude of $a^l = 1$.

The D²NN has several diffraction layers, with lightwave as input. In the output layer, photodetectors are classified. The D²NN is trained to focus light on the detector corresponding to the input beam.

### 3.1 Black-box optimization

Hyperparameter optimization is essential to improve the performance of neural networks. The hyperparameters of the D²NN in this study include the number of layers, distance between each layer, and selection/combination of the beam pattern. These are difficult to optimize using stochastic gradient methods, such as Adam optimizer [17] used in machine learning because finding the gradient calculation using an error back-propagation algorithm is difficult. Thus, we need other optimization methods to find the optimal hyperparameters.

The relationship between hyperparameter values and classification accuracy is unknown. Therefore, the best hyperparameter settings are obtained by training with various values and comparing losses. For this, grid and random searches have been used. Grid search is intuitive and can be parallelized. However, it evaluates nondominant hyperparameters, which increase the computation time [18]. Random search has the advantage of the search being terminated in the middle of optimization, and the results of multiple searches executed independently can be combined [19]. However, methods such as random and grid searches are inefficient because they require several computationally expensive learning calculations.

Therfore, we used black-box optimization. Black-box optimizations observe some hyperparameters $x$ and the value of the loss function $y$ and predict the $x$ value that minimizes $y$ from the results. Since the search proceeds while predicting the optimal hyperparameters, it is more efficient than random and grid searches.

TPE is a black-box optimization technique used as an automatic hyperparameter tuning algorithm for neural networks. TPE is a type of Bayesian optimization, consisting of a surrogate model to approximate the distribution of promising hyperparameters and an acquisition function to sample the next hyperparameters to be evaluated. TPE uses a proxy model and expected improvement as the acquisition function [20]. TPE first sets a random hyperparameter $x$ and evaluates it. After several evaluations, the evaluated hyperparameters are categorized into good and bad groups, and their distributions $l(x)$ and $g(x)$ are approximated by kernel density estimation. Next, several points are sampled, and $x$ is selected as the next observation point, maximizing $g(x)/l(x)$. By repeating this search, TPE can efficiently obtain excellent hyperparameters with few trials.

## 4. Results

In this study, we set the wavelength of HG beams to 1550 [nm], the beam waist to 0.01 [m], and the propagation distance through the atmosphere to 250 [m]. For the atmospheric disturbance, the internal scale $l_0 = 0.001$ [m], external scale $L_0 = 20$ [m], and atmospheric structure constant $C_n^2 = 1 \times 10^{-14}$ [m$^{-2/3}$] are assumed for free-space optical communication during calm daytime. For the D$^2$NN model, the number of pixels of the input beam plane, detector plane, and all layers is $256 \times 256$ pixels, and the sampling pitch is 39 [µm]. We prepared 1000 images for each beam mode, 800 images for training data, and 200 images for validation data. We fixed the distance from the input plane to the output plane at 30 [cm] and searched for the best layer arrangement. The loss function $E$ is the mean squared error:

$$E = \frac{1}{K}\sum_k (s_k - g_k)^2, \qquad (9)$$

where $K$ represents the number of pixels in the output image and ground-truth data, $s_k$ represents the light intensity of each pixel in the output image, and $g_k$ represents the light intensity of each pixel in the ground-truth image. For the ground-truth image, we prepared an image, in which only the detector region corresponding to each label was brightened. An example of our dataset is shown in Fig. 3, and parameters used for HG beam classification are listed in Table 1. In this study, we set the number of layers to two to improve the flexibility of tuning the interlayer distance.

Table 1:Parameters of the diffractive deep neural network model.

| Number of beam classifications ($m \times n$) | $(m, n)$ | Layers | Number of searches by TPE | Epochs | Batch size |
|---|---|---|---|---|---|
| 16 | (0,0)–(3,3) | 2 | 40 | 20 | 64 |
| 25 | (0,0)–(4,4) | | 40 | | |
| 36 | (0,0)–(5,5) | | 20 | | |

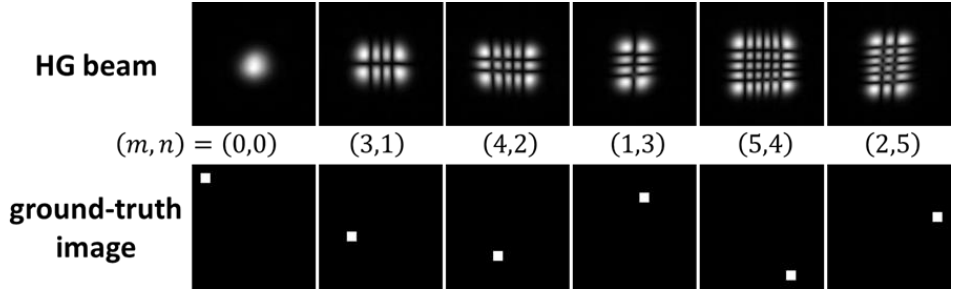

Figure 3:Example dataset: The top images are Hermite–Gaussian (HG) beams inputted to the diffractive deep neural network, and the bottom images are the corresponding ground-truth images. ($m$, n) represents the orders of *m* and *n* for an HG beam, respectively.

Table 2 shows the locations of layers and classification accuracy obtained using the proposed method. Table 2 also indicates the classification accuracy of the conventional method with equally spaced layers of 10 [cm], as shown in Fig.2 (a). Since classification accuracy depends on the randomness of the initial hyperparameter values of each layer, we trained the conventional method 40 times. We set the number of searches to 20 times in 36 classification modes because it takes a long time to train.

Table 2:Classification accuracy and the location of layers.

| Number of beam classifications ($m \times n$) | Maximum accuracy by conventional methods | Maximum accuracy by proposed method | $z_0$ [m] | $z_1$ [m] | $z_2$ [m] |
|---|---|---|---|---|---|
| 16 | 98.3% | 98.8% | 0.058 | 0.223 | 0.019 |
| 25 | 92.3% | 95.7% | 0.096 | 0.189 | 0.015 |
| 36 | 84.8% | 94.9% | 0.048 | 0.220 | 0.032 |

As shown in Table 2, comparing accuracy for 16 classification modes and auto-tuning the interlayer distance using TPE improved accuracy by 0.5% over equally spaced layer placement. Similarly, accuracy was improved by 3.4% for 25 classification modes and 10.1% for 36 classification modes over equally spaced layer placement. The difference in accuracy between TPE and equidistant placement is larger for 25 and 36 classification modes than for 16 classification modes. Note that the importance of auto-tuning the interlayer distance using

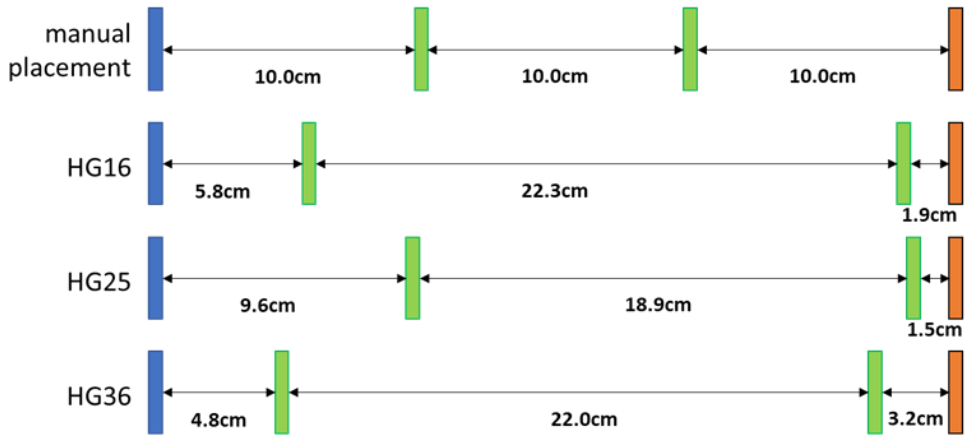

Figure 4：Distance between learned layers.

TPE increases as the number of modes increases. Figure 4 illustrates the interlayer distances obtained by TPE optimization. It is difficult to search such interlayer distances manually.

Figure 5 shows an example of an inferred image from the trained $D^2NN$. The inference was performed using the $D^2NN$ that achieved the highest accuracy of 94.9% in classifying the 36 modes of HG beams in Table 2. As shown in the figure, a specific detector region is brightened by the inference calculation. This light intensity can be used to determine original bits encoded in an HG beam using a photodetector. The surrounding detector areas become brighter along with the desired detector area. This results in a classification error. The error occurs in areas above, below, and to the left and right of the desired detector area. This error can be caused by the proximity of the detector areas or by the similarity of beam patterns due to the proximity of input beam modes. To confirm this, we randomly switched the detector's positions and the mode of each beam; however, the same errors occurred for beams with similar modes. For HG beam classification using the $D^2NN$, classification errors occurred between close modes and did not depend on the position of the detector.

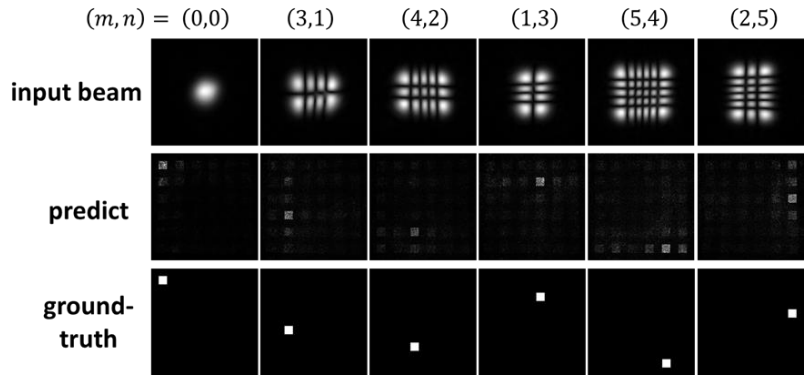

Figure 5：Example of inference results.

Next, we describe the computational time required for training. We used Windows 10 Enterprise as the operating system, AMD Ryzen 5 2600 as the CPU, and NVIDIA GeForce RTX 2060 as the GPU. We used TensorFlow-GPU 2.2.0 and Keras 2.4.3 as the deep learning framework. In this study, the batch size was 64, and the number of epochs was 20. The computation time for training was 650 ms per batch; approximately 43 min were required to train 16 modes and approximately 29 h to search 40 D²NN models. For the 25-mode classification, approximately 45 h were required to search 40 D²NN models and for the 36-mode classification, approximately 32 h were required to search 20 D²NN models.

## 5. Conclusion

In this study, we investigated how hyperparameter tuning could improve classification accuracy in multibit optical free-space communications with HG beams, when using a D²NN as a classifier. Results indicate that, by hyperparameter tuning, we can search for interlayer distances that yield higher classification accuracy than that obtained by manually placing D²NN layers at equal intervals. Classification accuracy was improved by approximately 0.5%, 3.4%, and 10.1% for 16, 25, and 36 classification modes, respectively, over equally placed spaced layers. In general, as the number of classification modes increases, the classification accuracy decreases, making it more important to improve accuracy by tuning the interlayer distance. To increase communication capacity in the future, it is necessary to increase the number of beam modes used, such as 49 or 64 modes. Automatic hyperparameter tuning should be performed to cope with an increase in communication capacity.

**Funding.** This research was funded by Japan Society for the Promotion of Science (19H04132,19H01097).

**Disclosures.** The authors declare no conflicts of interest.

**Data availability.** Data underlying the results presented in this paper are not publicly available but may be obtained from the authors upon reasonable request.

**Funding.** This research was funded by Japan Society for the Promotion of Science (19H04132,19H01097).

**Disclosures.** The authors declare no conflicts of interest.

**Data availability.** Data underlying the results presented in this paper are not publicly available but may be obtained from the authors upon reasonable request.

**References**

1. F. Pampaloni and J. Enderlein, "Gaussian, Hermite-Gaussian, and Laguerre-Gaussian beams: A primer," arXiv:physics/0410021 (2004).
2. G. Gibson, J. Courtial, M. Padgett, M. Vasnetsov, V. Pas'Ko, S. Barnett and S. F. Arnold, "Free-space information transfer using light beams carrying orbital angular momentum," Opt. Express **12**, 5448-5456 (2004).
3. L. R. Hofer, L. W. Jones, J. L. Goedert and R. V. Dragone, "Hermite–Gaussian mode detection via convolution neural networks," J. Opt. Soc. Am. A **36**, 936-943 (2019).
4. S. Fu, Y. Zhai, C. Yin, H. Zhou and C. Gao, "Mixed orbital angular momentum amplitude shift keying through a single hologram," OSA Continuum **1**, 295-309 (2018).
5. S. Jiang, H. Chi, X. Yu, S. Zheng X. Jin and X. Zhang, "Coherently demodulated orbital angular momentum shift keying system using a CNN-based image identifier as demodulator," Opt. Commun. **435**, 367-373 (2019).
6. Q. Tian, Z. Li, K. Hu, L. Zhu, X. Pan, Q. Zhang, Y. Wang, F. Tian, X. Yin and X. Xin, "Turbo-coded 16-ary OAM shift keying FSO communication system combining the CNN-based adaptive demodulator," Opt. Express **26**, 27849-27864 (2018).
7. Y. Zhou, J. Zhao, Z. Shi, S. M. H. Rafsanjani, M. Mirhosseini, Z. Zhu, A. E. Willer and R. W. Boyd, "Hermite–Gaussian mode sorter," Opt. Lett. **43**, 5263-5266 (2018).
8. M. Chagnon, M. Osman, Q. Zhuge, X. Xu and D. V. Plant, "Analysis and experimental demonstration of novel 8PolSK-QPSK modulation at 5 bits/symbol for passive mitigation of nonlinear impairments," Opt. Express **21**, 30204-30220 (2013).
9. T. Doster and A. T. Watnik, "Adaptive Demodulator Using Machine Learning for Orbital Angular Momentum Shift Keying," IEEE Photonics Technol. Lett. **29**, 1455-1458 (2017).
10. X. Lin, Y. Rivenson, N.T. Yardimci, M. Veli, Y. Luo, M. Jarrahi and A. Ozcan, "All-optical machine learning using diffractive deep neural networks," Science **361**, 1004–1008 (2018).
11. O. Zhao, S. Ha., Y. Wang, L. Wang and C. Xu, "Orbital angular momentum detection based on diffractive deep neural network," Opt. Commun. **443**, 245-249 (2019).
12. Z. Huang, P. Wang, J. Liu, W. Xiong, Y. He, J. Xiao, H. Ye, Y. Li, S. Chen and D. Fan, "All-Optical Signal Processing of Vortex Beams with Diffractive Deep Neural Networks," Phys. Rev. Appl. **15**, 014037 (2021).
13. P. Wang, W. Xiong, Z. Huang, Y. He, J. Liu, H. Ye, J. Xiao, Y. Li, D. Fan and S. Chen, "Diffractive Deep Neural Network for Optical Orbital Angular Momentum Multiplexing and Demultiplexing," IEEE J. Sel. Top. Quantum Electron. **28**, 1-11 (2021).